\documentclass[times, 10pt,twocolumn]{article}
\usepackage{BigDataBench_ISCA_2013}
\usepackage{times}
\usepackage{multirow}
\usepackage{graphicx} 
\usepackage{epsfig}
\usepackage{url}
\usepackage{authblk}
\usepackage{balance}
\usepackage[above]{placeins}

\usepackage{balance}

\begin{document}

\title{\emph{BigDataBench}: a Big Data Benchmark Suite from  Web Search Engines}

\author[1]{Wanling Gao}
\author[1]{Yuqing Zhu}
\author[1]{Zhen Jia}
\author[1]{Chunjie Luo}
\author[1]{Lei Wang}
\author[2]{Zhiguo Li}
\author[1]{Jianfeng Zhan \thanks{The corresponding author is Jianfeng Zhan.}}
\author[2]{Yong Qi}
\author[3]{Yongqiang He}
\author[4]{Shiming Gong}
\author[5]{Xiaona Li}
\author[6]{Shujie Zhang}
\author[7]{Bizhu Qiu}
\affil[1]{Institute of Computing Technology, Chinese Academy of Sciences, \authorcr \{gaowanling, zhuyuqing, jiazhen, luochunjie, wanglei2011, zhanjianfeng\}@ict.ac.cn}
\affil[2]{Xi'an Jiaotong University}
\affil[3]{Facebook, heyongqiang@fb.com}
\affil[4]{SouGou, yakergong@gmail.com}
\affil[5]{Baidu, lixiaona@baidu.com}
\affil[6]{Huawei, shujie.zhang@huawei.com}
\affil[7]{Yahoo!, qiubz@yahoo-inc.com}

\maketitle
\thispagestyle{empty}

\begin{abstract}

This paper presents our joint research efforts on big data benchmarking with several industrial partners. Considering the complexity, diversity, workload churns, and rapid evolution of big data systems, we take an incremental approach
in big data benchmarking. For the first step, we pay attention to search engines, which are the most important domain in Internet services in terms of the number of page views and daily visitors.
However, search engine service providers treat data, applications, and web access logs as business confidentiality, which prevents us from building benchmarks.
To overcome those difficulties, with several industry partners, we widely investigated the open source solutions in search engines, and obtained the permission of using anonymous Web access logs. 
Moreover, with two years' great efforts, we created a sematic search engine named \emph{ProfSearch} (available from \url{http://prof.ict.ac.cn}). These efforts pave the path for our big data benchmark suite from search engines---\emph{BigDataBench}, which is released on the web page (\url{http://prof.ict.ac.cn/BigDataBench}).

We report our detailed analysis of search engine workloads, and present our benchmarking methodology. An innovative data generation methodology and tool are proposed to generate scalable volumes of big data from a small seed of real data, preserving semantics and locality of data.
Also, we preliminarily report two case studies using \emph{BigDataBench} for both system and architecture researches.
\end{abstract}

\section{Introduction}\vspace{-12pt}

In recent years, more and more data are produced. The roles of people change from passive receptors of information to active creators. People are producing and sharing data continuously. It is reported that 2.5 quintillion bytes of data are created everyday \cite{dataproduce}. According to the survey of the global output by IDC, the data are growing exponentially now and the trends will be maintained in the coming years.
Big Data are considered as the asset of companies, organizations and even countries. Extracting the big value from Big Data requires enabling big data systems.

As researchers in both academia and industry pay great attention to innovative big data systems and architecture \cite{barroso2009datacenter} \cite{zhan2012high} \cite{ferdman2011clearing} \cite{ghazala}  \cite{lotfi2012scale}, the pressure of evaluating  and
comparing performance, energy efficiency, and cost-effectiveness of these systems rises \cite{zhancost} \cite{wang2012cloud}. 
Big data benchmarks are the cornerstone of those efforts.


In a tutorial given at HPCA 2013 \cite{HPCA_2013} \cite{scaledata}, we state our position view in big data benchmarking: we should take an incremental and iterative approach in stead of a top-down approach because of the following four reasons: first, there are many classes of big applications with a  lack of a scientific classification.
 Second, even for Internet service workloads, there are many important application domains, e.g., search engines, social networks, and electronic commerce, though they are mature in terms of both business and technology, however, customers, vendors, or researchers from academia or even different domains of industry  do not know enough to make a big data benchmark suite because of the confidential issues. Third, the value of big data drives the emergence of innovative application domains.  Fourth, the complexity,
diversity, workload churns, and rapid evolution of big data systems indicate that both customers and vendors often have incorrect or
outdated assumptions about workload behavior \cite{bigdatachen}.

This paper presents our joint research efforts on big data benchmarking with several industrial partners.
After investigating different application domains of Internet services---an important class of big data applications,
 we pay attention to search engines, which are the most important domain in Internet services in terms of the number of page views and daily visitors.
(Social networks, and electronic commerce follow, respectively.)
We widely investigated the open source solutions in search engines.
Moreover, with two years' great efforts, we created a sematic search engine named \emph{ProfSearch} (available from \url{http://prof.ict.ac.cn)}. These efforts pave the path for our big data benchmark suite from search engines---\emph{BigDataBench}.


We report our detailed analysis of search engine workloads, and present our benchmarking methodology. An innovative data generation methodology and tool are proposed to generate scalable volumes of big data from a small seed of real data, preserving semantics and locality of data.
Also, we preliminarily report two case studies using BigDataBench for both system and architecture researches. We gained two insights from the observations:
first,  the peak data processing rates of big data systems  are both applications and data volumes dependent, and hence tuning peak performance must consider different application scenarios.  Second, some architectural events, e. g., cache and TLB behaviors, are tending towards stability only on condition that the data volume increases to a certain extent. This observation has a significant implication for simulation-based architecture researches since large-scale simulation is time-consuming.

The rest of the paper is organized as follows. Section 2 presents the detailed analysis of search engine workloads. Section 3  summarizes our benchmarking methodology and and its result---\emph{BigDataBench}. Section 4 introduces how to generate scalable volumes of data. In Section 5, we report two case studies using \emph{BigDataBench} for both system and architecture researches. Finally, we draw the conclusion in Section 6.

\section{Workload analysis of search engines}\vspace{-12pt}

Our big data benchmarking work is  based on two practises: investigation of the open source solutions in common search engines, backed by several industry partners,
 and our experience of building a sematic search engine named \emph{ProfSearch}.

ProfSearch is a Chinese semantics search engine used to search scientists or professionals from different disciplines. Now it has collected publicly available information of 251,564 researchers across 260 universities and institutes. Different from common web search engines, for a scholar, ProfSearch can extract different categories of information from many data sources, for example, his/her research interests and educational background.

We have used various state-of-art algorithms or techniques to access, analyze, store big data and offer search services as follows \cite{jiacharacterization}:

\emph{Crawler workloads}. We use Scrapy \cite{scrapy}, a widely used open source web scraping framework written in Python, to build our web crawler.

\emph{Analysis workloads}. We have used various machine learning and data mining techniques to classify and cluster web pages downloaded by \emph{Crawler}, extract structured data from unstructured web pages, analyze the semantics and topics of these pages. These analysis workloads contain state-of-art algorithms and learning models, such as Support Vector Machine (SVM) \cite{cortes1995support}, Naive Bayes, K-means, Hidden Markov Model (HMM) \cite{rabiner1989tutorial}, Conditional Random Fields (CRFs) \cite{lafferty2001conditional}, Latent Semantic Analysis (LSA) \cite{deerwester1990indexing} and Latent Dirichlet Allocation (LDA) \cite{blei2003latent}.

SVM and Naive Bayes are widely used classification algorithms in data mining. We use SVM to classify whether a web page is a home page or not, and Naive Bayes to classify professionals to a certain category according to his/her research interests.  Different from classification which is a supervised learning process, clustering is unsupervised without manual annotation. We use \emph{K-means} to cluster papers to find the similarity among scientists' papers.

HMM and CRFs are statistical sequence modeling method, often applied in speech, handwriting, gesture recognition, part-of-speech tagging, and bioinformatics. Whereas an ordinary classifier predicts a label for a single sample without regard to "neighboring" samples, these models can take context into account. We use HMM for Chinese segmentation which divides the text of Chinese into meaningful tokens and part-of-speech tagging, and CRFs for information extraction which automatically extracts structured data from web pages.

LSA uses a mathematical technique called singular value decomposition (SVD) to identify patterns in the relationships between the terms and concepts contained in an unstructured collection of text. It uncovers the underlying latent semantic structure of word usage in a body of text.  LDA is an example of a topic model. It is a generative model that allows sets of observations to be explained by unobserved groups that explain why some parts of the data are similar. We have used LSA and LDA to analyze what research fields are similar to other's interests and who has similar interests with the others.

\emph{Store and management workloads}. In ProfSearch, there is large scale of unstructured, semi-structured and structured data to be stored. The original unstructured web pages are stored using HDFS, the distributed file system in Hadoop \cite{hadoop}. Large intermediate data generated in the analysis process are semi-structured, and we use Hive \cite{hive} to store them for further analysis. The structured data extracted from the web are stored in MySQL.

\emph{Web service workloads}. To offer web search service, we use Sphnix \cite{sphinxsearch}, a wildly used free software, to index documents and provide results for query. Sphnix implements inverted index. It is an index data structure storing a mapping from content to its locations in a database file, or in a document or a set of documents. Sphnix uses slightly modified BM25 function to rank matching documents according to their relevance to a given search query. BM25 is based on the probabilistic retrieval framework and represents state-of-art TF-IDF-like retrieval functions. Additionally, we use Apache Tomcat as the web server, and Memcached \cite{memcached}, a distributed memory object caching system, to speed up the search service.\vspace{-6pt}

\section{Benchmarking Methodology and Decisions}\vspace{-12pt}

This section presents our methodology and decisions on assembling \emph{BigDataBench}. We state how we choose the representative applications and generate scalable volumes of  data preserving semantics and locality of data. And we also give a summary of \emph{BigDataBench}.\vspace{-12pt}


\subsection{Following An Incremental Approach}\vspace{-12pt}

For the first step, we investigate application domains. We single out the most important application domain. Firstly, we pay attention to Internet services, and  rank main application domains according to a widely acceptable metric ---the number of page views and daily visitors.
We investigate the top sites listed in Alexa \cite{Alexa}, of which the rank of the sites is calculated using a combination of average daily visitors and page views. We find the search engine is the mostly popular application domain, 40\% of the top sites are search engines, such as Google, Bing and etc. Data from the  Internet study \cite{pew} also prove the popularity of search engines, which shows that 92\% of online adults use search engines to find information on the web. So we focus on  the search engine firstly.


For the second step, we choose typical workloads from Web search engines  as candidates of  our BigDataBench.

 Considering the workload churns and emerging workloads domains, we believe that the mature big data benchmark suites will take a long way to go.
We will continuously add more representative applications and remove the out-of-data applications.
\vspace{-12pt}


\subsection{Methodology of Generating Big Data}\vspace{-12pt}

 Big data benchmarks require big data set as the inputs to drive their workloads.
 Chen et al.\cite{bigdatachen} found that the observed workload behaviors, shown in their collected large-scale, long-term MapReduce workloads traces from Cloudera and Facebook, do not fit well-known statistical distributions. Our previous work \cite{xi2011characterization} collected three search engine companies' traces and also found that  the frequently used distributions cannot capture the key characteristics of real data.
 Moreover, there is a large performance gap between running search engines with real data and randomly generated data \cite{xi2011characterization}.
 In this case, only real data can reflect the real system behaviors, and hence the real life data is preferred \cite{bigdatachen} in big data benchmarks.
 However, it is a big challenge to obtain real big data as follows: 
Firstly, most of  end user do not own real big data whereas
Internet service companies who own the real life big data would not like to share big data for commercial
confidentiality and user privacies; secondly, even though big data is openly available, downloading terabyte scale data is too costly to be acceptable.



Based on the above reasons, we would like to generate synthetic data preserving key characteristics of real data.
There are two key characteristics we must consider semantic and locality. Semantic characteristic reflects the insightful meaning of real life data. Locality reflects the data access patterns. We investigate the real life data with the purpose of getting a semantic model and a locality model. We find the real search engine query terms follow zipf's law\cite{xi2011characterization}. We then generate the synthetic query trace on the basis of the real search query terms we have gotten and let the
query terms follow zipf's law. The temporal locality can be reflected by using reuse distance. We calculate each real term's reuse distance and generate the synthetic data according to the real terms' reuse distance.  The details can be found at Section 4.\vspace{-12pt}


\subsection{Considering Variety of Workloads}\vspace{-12pt}

In addition to three "V" of big data: volumes, velocity, and variety, our previous work \cite{scaledata} showed that diversity of workloads must be considered in big data benchmarking since they have different characteristics in term of computation, memory, and I/O access patterns. As we discussed in Section 2, a search engine involves in many important workloads, and we must choose typical workload for assembling BigDataBench.
\vspace{-12pt}


\subsection{Summary of \emph{BigDataBench}}\vspace{-12pt}



 After the analysis of the common search engines backed by several industry partners and our semantics search engine-ProfSearch, we choose the following workloads for \emph{BigDataBench} at present: \emph{Sort, Grep, WordCount, Naive Bayes } and \emph{SVM} for their representative algorithms and diversities of computing and I/O access patterns \cite{chenzheng}. We also choose the search service including a back-end search server and a front-end Web server. Our future work will add more workloads.\vspace{-12pt}

\section{Scalable Data Generation  Tool}\vspace{-12pt}

The data generation implementation of BigDataBench includes two parts. One is generating user requests, and the other is generating input data of BigDataBench. \vspace{-12pt}

\subsection{Request Generation \cite{xi2011characterization}} \vspace{-6pt}


The key characteristics of a search workload
trace are query sequences and timing sequences \cite{xi2011characterization}. Query
sequences depict the contents in each request. Query contents
are determined by the semantic model that characterizes
the frequencies of terms and the combinations of terms
constituting a request. The timing sequences depict the issuing
intervals of requests, from which we can compute
the fluctuation of query requests. Meanwhile, measuring the temporal locality of a series of queries
is similar with measuring the temporal locality of a series
of memory accesses. Stack distance  is an effective
way to analyze the temporal locality of a series of requests.
Stack distance, also called reuse distance, depicts the number
of different queries between a recurring query.  we have obtained
permission to use three real workload traces, one from SoGou  and the other two from two of the largest search
service providers in China.

We  generate synthetic query rate according to a real query trace. Supposing that we have an one day real-life trace, we divide the whole trace to several sections by one hour or one minute so that it can reflect the user's behaviors more accurately.
So we generate scalable volumes of request traces preserving timing, semantics, and locality model. The details of those models  can be found at our previous work \cite{xi2011characterization}.

\subsection{Input Data Generation}\vspace{-6pt}

Our data generation tool is based on small-scale real data. We firstly analyze the characteristics of the small-scale data and then expand the data scale maintaining the characteristics. To preserve the characteristics of real-world data, we analyze the real data from semantic and locality.

We parse the real data and get class information, word information and the length of documents. Class information includes all the classes that occurred, the number of documents in each class, and the number of all the documents. Word information includes all the words that occurred, the number of a word occurred in each class, and the number of words in each class. Basing on those information, we can compute the probability distribution of classes, the probability distribution of words in each class, and the probability distribution of document length in each class. We expand the real data to big data by keeping those probability distributions unchanged.\vspace{-12pt}


\section{Case Studies}\vspace{-12pt}
In order to evaluate \emph{BigDataBench}, we do some case studies in this section. First, the data generation is the most important part of Big Data Benchmarking, so we compare our generated data with real data to evaluate the validity of our generated data. Second, we present two case studies running \emph{BigDataBench} for system and architecture researches respectively.

In the first experiment, we deploy a distribution of Hadoop \cite{hadoop} over a 2-node cluster (one master and one slave node). In the latter two experiments, we deploy a distribution of Hadoop over a 15-node cluster (one master and fourteen slave nodes). All the nodes in the clusters have the same configuration. Each node has two Xeon E5645 processors equipped with 16 GB memory and 8 TB disk. The detailed configuration for each node is listed in Table 2. The operating system is Centos 5.5 with Linux kernel 2.6.34. The Hadoop distribution is 1.0.2 with Java version of 1.6. The configuration of Hadoop is 18 map slots and 18 reduce slots for each slave node, and the Java opts for the task tracker child processes are 512 MB. In order to make the results credible, we repeat each experiment three times and report the average value.

\begin{table}[h]
\center
\begin{tabular}{|c|c|c|c|}
  \hline
  \multicolumn{2}{|c|}{CPU Type} & \multicolumn{2}{|c|}{Intel CPU Core} \\ \hline
  \multicolumn{2}{|c|}{Intel \textregistered Xeon E5645}  &\multicolumn{2}{|c|}{6 cores@2.40G} \\ \hline
  \hline
L1 DCache &L1 ICache &L2 Cache &L3 Cache \\ \hline
6 $\times$ 32 KB& 6 $\times$ 32 KB&6 $\times$ 256 KB& 12MB \\ \hline
\end{tabular}
\caption{Details of node configuration}\label{hwconfigeration}\vspace{-6pt}
\end{table}

\subsection{Comparing Generated Data with Real Data}\vspace{-6pt}
To evaluate the correctness of our methodology, we conduct an experiment to compare the \emph{BigDataBench}'s generated data with the real data running five \emph{BigDataBench} workloads. The  real data is a 146-MB wikipedia data while the 146-MB synthetic data is generated by our tool using 7MB of the wikipedia as the seed. The selected five workloads
are Sort application, Grep application, Wordcount application, Naive Bayes application and SVM application from \emph{BigDataBench}.
We collect application-level and architectural-level monitor statistics when running the workloads.
We use a user-perceived performance metric---data processing rate to evaluate the system processing capability \cite{luo2012cloudrank} under different data set. For each application, the metric of data processing rate is defined as the input data size dividing by the application running time. For example, the running time of Sort with 100GB input is 2487 second, and  then the data processed per second of Sort at 100GB is 41.6MB/s.
Architectural statistics are collected by using the \emph{perf} tool. We investigate the following metric: L3 cache misses per 1000 instructions, L2 cache misses per 1000 instructions, L1 Instruction misses per 1000 instructions, L1 data misses per 1000 instructions, Instruction TLB (Translation Lookaside Buffer ) misses per 1000 instructions and  data TLB misses per 1000 instructions. We have not got the value of  L1 data cache misses as it can be overlapped by out-of-order executions.

\begin{figure}[htb]
\centering
\includegraphics[scale=0.62]{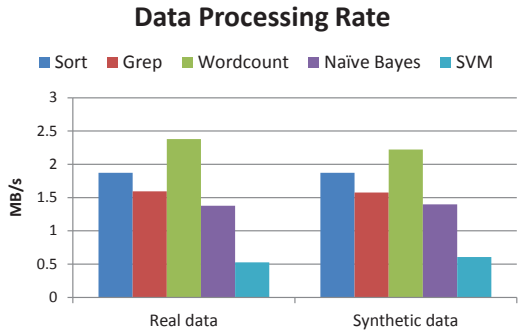}
\caption{The Data Processing Rates of Real Data and Synthetic Data.} 
\label{wiki-cmp-process}
\end{figure}

\begin{figure*}[htb]
   \centering
   \includegraphics[scale=0.68,angle=270]{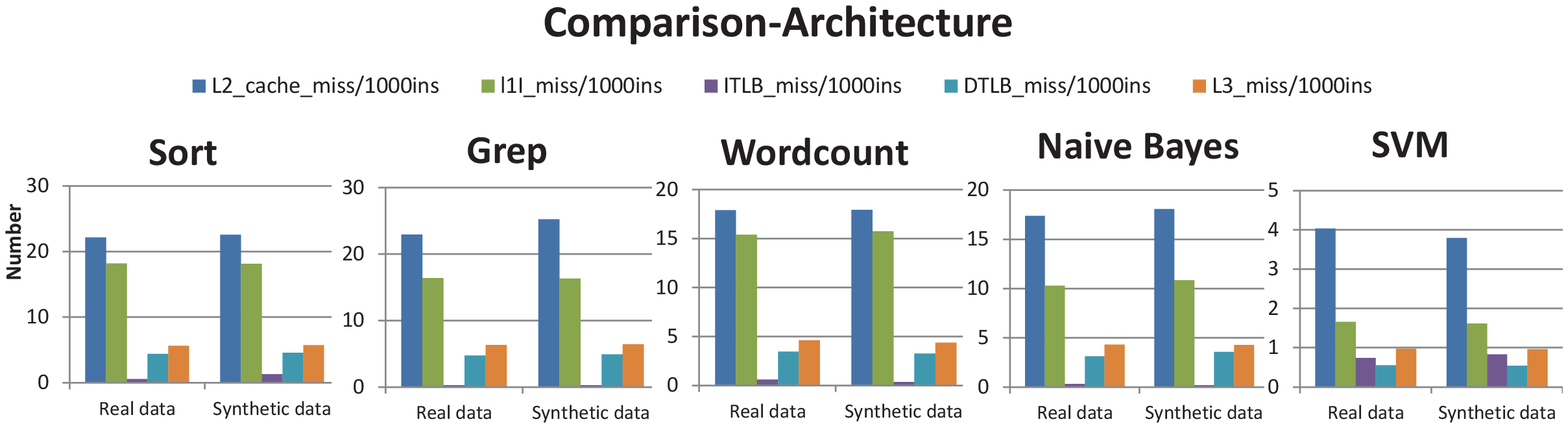}
   \caption{Cache and TLB Behaviors of Real Data and Synthetic Data.}
   \label{wiki-cmp-arch}
\end{figure*}

Figure \ref{wiki-cmp-process}, Figure \ref{wiki-cmp-arch} separately reports the data processing rates, cache and TLB behaviors of running two kinds of data with regard to different workloads. From Figure 1, we can see that for the synthetic and real data, the data processing rates of the workloads are close and the deviation of two data sets with the same workload is less than 12.9\% (SVM), and the deviation of the two data sets are about 1\% for Sort, Grep, Naive Bayes and 7\% for Wordcount, which implied that the data processing behaviour is consistent between the generated data and the real data. From Figure 2, we can see that the cache and TLB behaviors for the real and synthetic data are close and the deviation of two data sets with the same workload is less than 2 instructions per 1000 instructions, which implied that the memory access behaviour is consistent between the generated data and the real data. So we can say that for the system or architecture research using the generated data of \emph{BigDataBench} can achieve the similar workload behaviours with that of the real data.

\subsection{Using \emph{BigDataBench} for System Evaluation}\vspace{-8pt}

In this section, we use \emph{BigDataBench} to evaluate the performance of the cluster system mentioned above.
As Rajaraman explained \cite{rajaraman2008more}, for big data application, inferior algorithms beat better than sophisticated algorithms because of the computing overhead.  We choose five applications that use simple algorithms from \emph{BigDataBench}: \emph{WordCount}, \emph{Naive Bayes}, \emph{Grep}, \emph{SVM} and \emph{Sort} whose computation complexities slightly vary from $O(n)$ to  $O(n*log_2n)$. Similar with the above case study, we still use data processing rate to evaluate the system processing capability.



\begin{figure}[htb]
\centering
\includegraphics[scale=0.8,width=3.45in,height=2.5in]{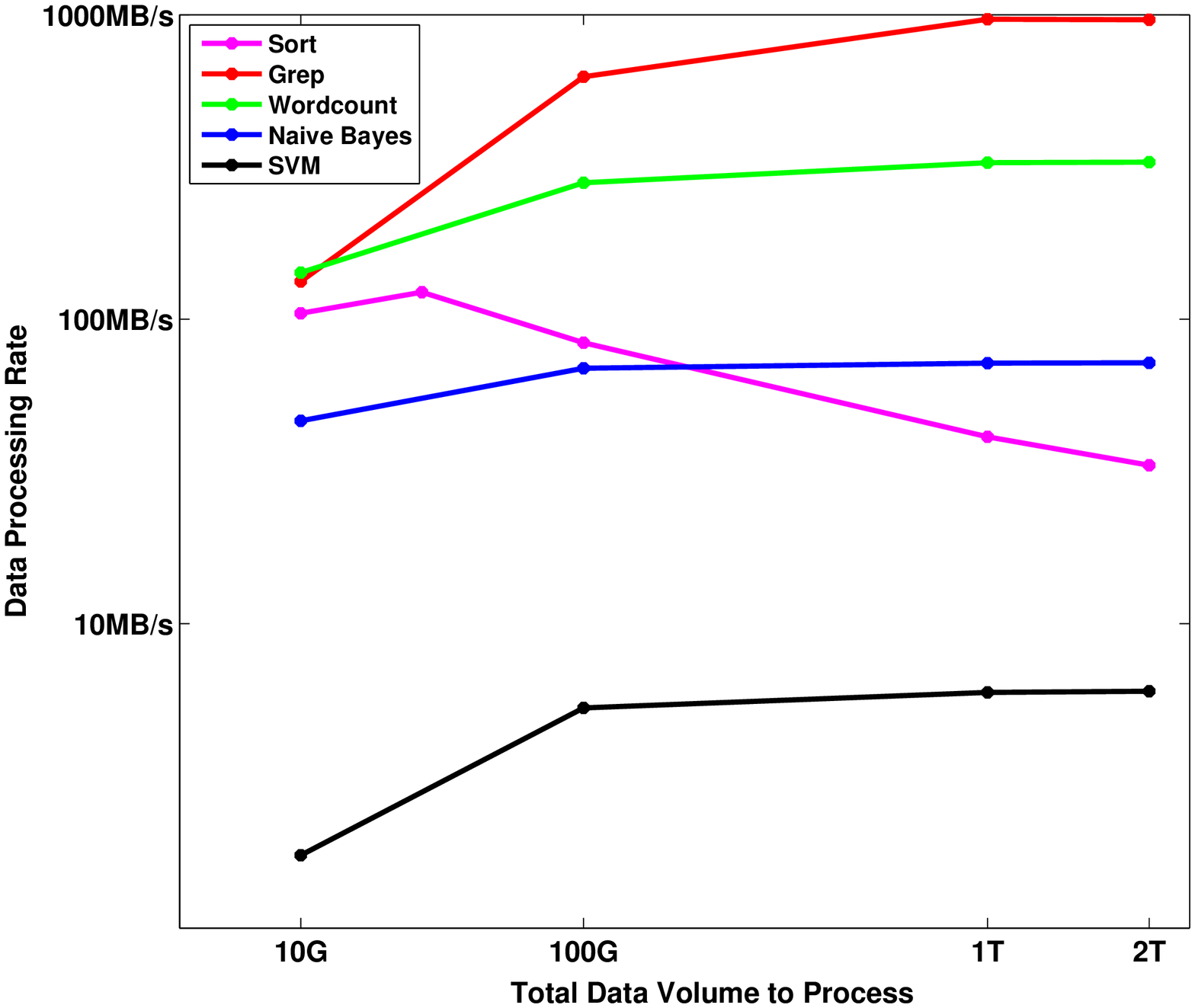}
\caption{Data Processing Rates for Different Data Volumes to Process.}
\label{unitdata}
\end{figure}

Figure \ref{unitdata} reports the data processing rates for different data volumes with regard to different workloads. Please note that the same data is employed for all five workloads. Therefore, as shown in the figure,  the system under test have a threshold with regards to data processing rate for \emph{Grep, WordCount, Naive Bayes and SVM } workloads. The cluster system can only be fully loaded on condition that the data volume exceeds the threshold.
From the figure, we can observe that the thresholds for the four workloads are between 100GB and 1TB. 
In comparison, \emph{Sort} has global data access requirements. From Figure \ref{unitdata}, we can find that there exist an inflexion point. After this point, the data processing rate decreases. Thus, there must exist a data size on which the data processing rate is optimal for the cluster system and the workload. This data size should fall between 10GB and 1TB. The reason for the existence of the inflexion point is that Sort requires the global transfer of the whole data set. Besides, data must be transferred for processing. Thus, after the inflexion point, the data processing waits for the data transfer, which can possibly saturate the I/O and the network. The larger the data volume, the more data to be transferred. Given the same bandwidth and I/O throughput, the longer time is needed for data transfer, thus the longer the data processing must wait for the data transfer. This finally results in the slope of the curve in Figure \ref{unitdata} for the sort workload.

Different data volume thresholds  and inflexion point (for some workloads) are found in different workloads of close computation complexity. Meanwhile, the peak data processing rates
have large performance gaps among different applications. These are the key characteristics of \emph{BigDataBench}, indicating so-called peak performance are both application and data volume dependent.

\vspace{-8pt}

\subsection{Using \emph{BigDataBench} for Architecture Research}\vspace{-8pt}

We also collect the micro-architectural statistics to analyze the architectural metrics on experiments with different data scales. Architectural statistics are collected by using the \emph{perf} tool.  We analyze the micro-architecture from the perspective of
cache access efficiency.  There is a complex memory hierarchy  for modern processor. So we investigate the following metric: L3 cache misses per 1000 instructions, L2 cache misses per 1000 instructions, L1 Instruction misses per 1000 instructions, instruction TLB (Translation Lookaside Buffer ) misses per 1000 instructions and  dada TLB misses per 1000 instructions. We have not got the value of  L1 data cache misses as it can be overlapped by out-of-order executions.
Thus the chosen metrics basically cover the  complete memory hierarchy of the  processor.



\begin{figure*}[htb]
   \centering
   \includegraphics[scale=0.68,angle=270]{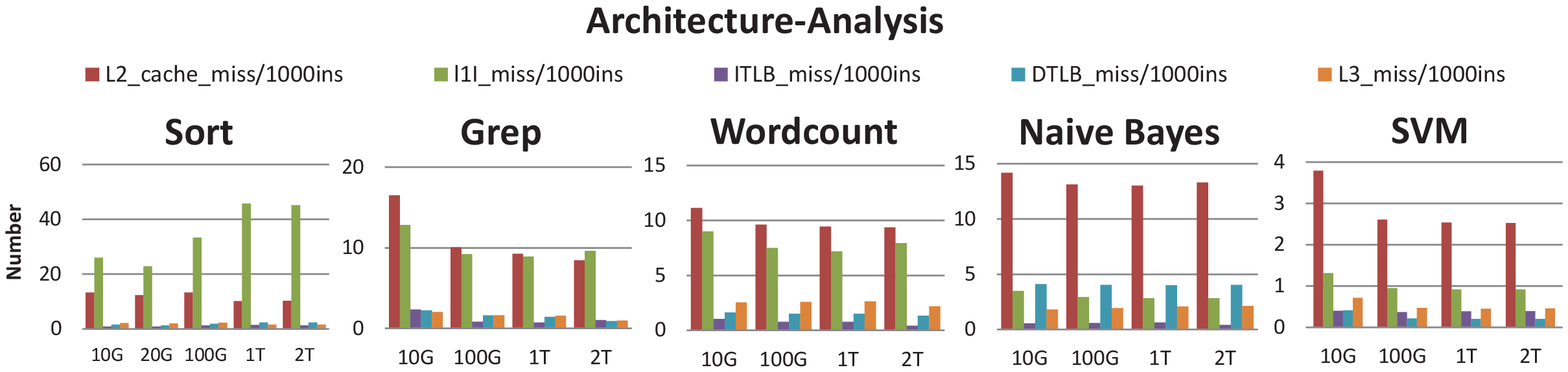}
   \caption{Cache and TLB Behaviors of Data Analysis Applications.}
   \label{arch}
\end{figure*}

\begin{figure}[htb]
\centering
\vspace{-12pt}
\includegraphics[scale=0.62]{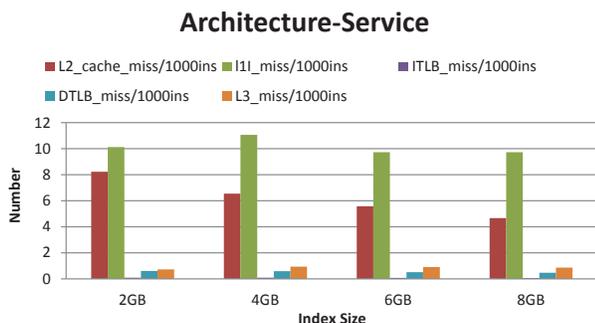}
\caption{Cache and TLB Behaviors of the Nutch Server.} 
\label{service}
\end{figure}

The cache and TLB behaviors of the system under five data analysis workloads from \emph{BigDataBench} workloads are shown in Figure \ref{arch}.
As a comparison, we also benchmark a search sever of search engine, which uses the distribution version \emph{Nutch 1.1}. The search sever receives user requests, searches in the index, and then sends the corresponding items to front end. In this benchmarking effort, we change the index size from 2 GB to 8 GB, and the corresponding segment size from 4.4 GB to 17.6 GB. The cache and TLB behaviors of the search server  are shown in Figure \ref{service}.


From Figure \ref{arch} and Figure \ref{service}, we can find that for different application, the cache and TLB behavior data have different trend with increasing data volumes. For example, for  L1 Instruction misses per 1000 instructions, the number of \emph{Sort} increases with the volume of data input while the number of \emph{Grep} decreases. However, for different (five) applications, the numbers reach stable values when the data volume increases to a certain extent.


In sum, we conclude that  some architectural events  are tending towards stability only on condition that the data volume increases to a certain extent. This observation has a significant implication for architecture researches since simulation is time-consuming.
We will perform further investigation.
\vspace{-8pt}

\section{Conclusions}\vspace{-8pt}

In this paper, we present our joint research efforts with several industrial partners on big data benchmarking from Web search engines:  the most important domain in Internet services in terms of the number of page views and daily visitors.

Our benchmarking methodology includes three parts: first, we follow an incremental approach to assemble big data benchmarks, and then an innovative data generation methodology and tool are proposed to generate scalable volumes of big data from a small seed of real data, preserving semantics and locality of data. Finally, we consider diversity of workloads in addition to volumes, velocity and variety of data.
Also, we preliminarily report an experiment to verify correctness of the data generation methodology of BigDataBench and two case studies using BigDataBench for both system and architecture researches. We gained two insights from the  observations:
first,  the peak data processing rates of big data systems  are both applications and data volumes dependent, and hence tuning peak performance must consider different application scenarios.  Second, some architectural events, e. g., cache and TLB behaviors, are tending towards stability only on condition that the data volume increases to a certain extent. This observation has a significant implication for simulation-based architecture researches since large-scale simulation is time-consuming.

\vspace{-8pt}

\section{Acknowledgements}\vspace{-8pt}
We are very grateful to anonymous reviewers. This work is supported by the Chinese 973 project (Grant No.2011CB302502), the Hi-Tech Research and Development (863) Program of China (Grant No.2011AA01A203, No.2013AA01A213), the NSFC project (Grant No.60933003, No.61202075) and the BNSF project (Grant No.4133081).\vspace{-8pt}

\balance
\nocite{ex1,ex2}

\end{document}